\documentclass[letterpaper, 10 pt, conference]{ieeeconf}

\usepackage{soul, xcolor}

\IEEEoverridecommandlockouts  
\usepackage[T1]{fontenc} 
\usepackage{amsmath,amssymb,amsfonts}
\usepackage{mathtools}
\usepackage{bm}
\usepackage{mathrsfs}
\usepackage{bbm}
\usepackage{graphicx}
\usepackage{array}
\usepackage{tabularx}
\usepackage{multirow}
\usepackage{multicol} 
\usepackage{subcaption} 
\usepackage{tikz,pgf}
\usetikzlibrary{calc,positioning,mindmap,trees,decorations.pathreplacing}
\usepackage{cite}
\usepackage{verbatim}

\newcounter{definition}
\newenvironment{definition}[1][]
{	
	\par\addvspace{0.5\baselineskip}
	\refstepcounter{definition}
	\noindent\textbf{Definition~\thedefinition%
		\if\relax\detokenize{#1}\relax\else\ (#1)\fi. }\normalfont
}
{
	\par\addvspace{0.5\baselineskip}
}

\newcounter{theorem}
\newenvironment{theorem}[1][]
{	
	\par\addvspace{0.5\baselineskip}
	\refstepcounter{theorem}
	\noindent\textbf{Theorem~\thetheorem%
		\if\relax\detokenize{#1}\relax\else\ (#1)\fi. }\normalfont
}
{
	\par\addvspace{0.5\baselineskip}
}

\newcounter{lemma}
\newenvironment{lemma}[1][]
{	
	\par\addvspace{0.5\baselineskip}
	\refstepcounter{lemma}
	\noindent\textbf{Lemma~\thelemma%
		\if\relax\detokenize{#1}\relax\else\ (#1)\fi. }\normalfont
}
{
	\par\addvspace{0.5\baselineskip}
}

\newcounter{assumption}
\newenvironment{assumption}[1][]
{	
	\par\addvspace{0.2\baselineskip}
	\refstepcounter{assumption}
	\noindent\textbf{Assumption~\theassumption%
		\if\relax\detokenize{#1}\relax\else\ (#1)\fi. }\normalfont
}
{
	\par\addvspace{0.2\baselineskip}
}

\newtheorem{remark}{Remark}

\providecommand{\keywords}[1]{\textbf{\textit{Index terms---}} #1}
\title{\LARGE \bf
	Exact Average Consensus under Noisy Communication Links: \\ A Decentralized Gradient Perspective
}

\author{Yuhang Deng, Zheng Chen, and Erik G. Larsson
	\thanks{This work was supported in part by ELLIIT, the Swedish Research Council (VR), and the Knut and Alice Wallenberg Foundation.}
	\thanks{The authors are with the Department of Electrical Engineering, Link\"oping University, 58183 Link\"oping, Sweden. 
		{\tt\small Email: \{yuhang.deng, zheng.chen, erik.g.larsson\}@liu.se}.}%
}

\begin{document}

\maketitle
\thispagestyle{empty}
\pagestyle{empty}

\begin{abstract}
    We study the distributed average consensus problem under persistent link-level disturbances modeled as a martingale difference sequence with uniformly bounded conditional second moments. 
    Under such disturbances, the standard stochastic-approximation-based linear iteration with diminishing stepsizes drives the network to consensus on an unbiased random variable with non-vanishing variance instead of the exact initial average.
    To understand and resolve this limitation, we develop an anchoring-based mechanism derived from a decentralized gradient descent formulation and study the effect of incorporating a decaying anchoring term that continuously pulls each agent state toward its initial value.  
    This perspective provides an intuitive interpretation of how state anchoring counteracts disturbance accumulation. Under standard summability conditions, we prove that the resulting algorithm achieves exact average consensus almost surely. 
    Furthermore, this decentralized gradient perspective offers a unifying framework for several related methods and an interpretable design principle for exact average consensus under persistent disturbances.
\end{abstract}

\section{INTRODUCTION}
Multi-agent cooperation and control is the building block of many engineering applications, particularly in large-scale networks consisting of multiple intelligent agents engaged in a common decision-making process \cite{4140748RenWei, 4118472ReZa}. In the absence of a central coordinator, agents must rely on local interactions to reach agreement on certain quantities. This challenge has driven significant interest in distributed consensus problems within the signal processing and control communities\cite{5545370Dimakis}. A notable special case is the average consensus problem, whose objective is for all agents to agree on the average of their initial state values. It underpins numerous applications such as distributed sensing and cooperative control, where the exact average serves as a global estimate or coordination reference, and even small deviations from it may propagate into subsequent decision-making.

One of the most classic approaches to achieving average consensus is the distributed linear iteration algorithm, in which each agent iteratively updates its state by forming a linear combination of its own value and the information received from its neighbors. Its canonical form is given by
\begin{equation}
	\label{Eq.LIA}
	\vspace{-1mm}
	x_i[t \!+\!1] = w_{ii}\, x_i[t] + \sum_{j \in \mathcal{N}_i} w_{ij} x_j[t], \quad \forall i \in \mathcal{N},
\end{equation}
where $x_i[t] \in \mathbb{R}$ represents the state value held by agent $i$ in iteration $t$, $\mathcal{N}=\{1,\ldots,N\}$ denotes the set of agents, $\mathcal{N}_i$ is the set of neighbors of agent $i$, and $w_{ij} \in \mathbb{R}$ is the mixing weight that agent $i$ assigns to information received from its neighbor agent $j$. The convergence properties of \eqref{Eq.LIA} depend on the mixing matrix $\mathbf{W}=[w_{ij}]_{i\in[N],j\in[N]}$, and have been extensively analyzed in \cite{200465Lin,olshevsky2009convergence}.

In practical networks, information transmission may be corrupted by noise or other impairments in the communication process. The received information from agent $j$ to agent $i$ under communication uncertainty can be modeled as\footnote{Here, we do not consider any disturbance on the self-link, since the self-state is available locally and is not communicated.}
\begin{equation}
\label{noisy_transmitted_signal}
y_{ij}[t] = x_j[t] + \epsilon_{ij}[t],
\end{equation}
where $\epsilon_{ij}[t]$ represents link-level disturbance, typically assumed to be zero-mean. This formulation also includes the case with quantization noise, in which the disturbance value $\epsilon_{ij}[t]$ is the same for all neighbors of agent $j$. The consensus update then becomes
\begin{equation}
\label{noisy_LTI}
x_i[t+1] = w_{ii}x_i[t]+\sum_{j \in \mathcal{N}_i} w_{ij} y_{ij}[t].
\end{equation}
Without any further modification, it is shown in \cite{XIAO200733} that the network average follows a random walk, which prevents the convergence to a fixed consensus value. 

Average consensus with imperfect communication has been extensively investigated in the literature.
Existing studies have considered various sources of disturbance, including channel fading \cite{HUANG20101571,XU2018503}, quantization errors \cite{5929538Lavaei,Nedic4738891} and additive link noise \cite{Huang06067359X, LI5411807}. 
To mitigate the effects of link-level disturbance, many prior works have adopted stochastic approximation (SA) methods with a sequence of decaying stepsizes. A general form of the update rule is
\begin{equation}
    \label{SA}
    x_i[t+1] = x_i[t] + \alpha[t] \sum_{j \in \mathcal{N}_i} w_{ij}\left(y_{ij}[t] - x_i[t]\right),
\end{equation}
where $\alpha[t]$ is the stepsize satisfying Assumption \ref{Assumption: stepsize}.
\begin{assumption}
\label{Assumption: stepsize}
The stepsize sequence $\{\alpha[t]\}_{t=0}^{\infty}$ is positive and
non-increasing, satisfying $0<\alpha[t] <1$ for all $t$, $\alpha[t] \to 0$, and the standard summability conditions: 
\begin{equation}
    \sum_{t=0}^{\infty} \alpha[t]=\infty, \quad  \sum_{t=0}^{\infty} \alpha[t]^2<\infty.
\end{equation}
\end{assumption}

SA-based consensus algorithms have been studied in \cite{Huang06067359X, Kar4663899,Touri5203882,LI20091929,LI5411807, Pescosolido4641610}, under various graph topologies, noise types, and convergence guarantees. 
A common conclusion is that these algorithms typically do not converge to the exact average of the initial state values; instead, they converge to a random variable whose expected value equals the exact average but with nonzero variance. To compensate for the random deviation in the consensus value, \cite{Kar4663899} applies Monte Carlo averaging at the cost of additional energy and runtime.

Along a different line of research, several studies have explored ways to achieve robust average consensus by incorporating the initial information into the iteration. In \cite{Pescosolido4641610},  a leakage-based mechanism is introduced to preserve the network average in expectation, but the disturbances still introduce bias in the agent's asymptotic states. For directed graphs, a similar idea has been explored \cite{Vivek10286415}, which augments the standard push sum algorithm to include a weighted initial state in the update rule.

Note that many noise-mitigating average consensus algorithms rely on heuristic rules and lack clear, intuitive interpretations of the design principle. We bridge this gap by viewing this problem from a decentralized gradient perspective that explains how incorporating initial state information mitigates persistent disturbances. Indeed, average consensus is a special case of distributed optimization   with a sum of quadratic functions as the global objective. Applying decentralized gradient descent (DGD) with decaying stepsizes can achieve almost sure convergence to the exact minimizer of the global objective, which is the exact initial average \cite{4749425Nedic}. This DGD-based formulation leads to an interesting iterative update rule that combines three terms in every iteration: 
\begin{itemize}
\item the agent’s current state value;
\item a weighted combination of the neighbors’ contributions, scaled by a time-decaying stepsize;
\item the agent’s initial state value, scaled by another time-decaying stepsize.
\end{itemize}
We show that with properly chosen sequences of decaying stepsizes, this algorithm achieves almost sure convergence to the exact average consensus value under link-level disturbances modeled as a Martingale Difference Sequence (MDS). Our framework offers an intuitive interpretation of existing algorithms under a similar design principle, including noise-resilient push-sum \cite{Vivek10286415}, dynamic average consensus \cite{Wang10383541}, and over-the-air distributed consensus \cite{Deng11161728}.

\section{Preliminaries on Average Consensus with Noisy Communication Links}

We consider a distributed network of $N$ agents represented by a time-invariant undirected graph $\mathcal{G}=(\mathcal{N},\mathcal{E})$, where $\mathcal{N}=\{1,\ldots,N\}$ is the set of agents and $\mathcal{E} \subseteq\mathcal{N}\times\mathcal{N}$ is the set of links. For any $i,j \in \mathcal{N}$, $(i,j) \in \mathcal{E}$ holds if agents $i$ and $j$ can communicate. Recall the distributed linear consensus iteration in an ideal environment with noise-free communication:
\begin{equation}
    \bm x[t+1]=\mathbf{W} \bm x[t].
    \label{eq:distributed-iteration}
\end{equation}
Here, $\bm x[t]=[x_1[t],\ldots, x_N[t]]^\top$ contains the state values of all agents in iteration $t$, and $\mathbf{W}$ contains the mixing weights, with $w_{ij}=0$ if $(i,j)\notin\mathcal{E}, i\neq j$. The following assumption on $\mathbf W$ is imposed throughout the paper.
\begin{assumption}
\label{assump.1}
The \textit{mixing matrix} $\mathbf{W}$ is symmetric and satisfies $\mathbf{W}\mathbf{1}=\mathbf{1}$, $\mathbf{1}^{\top} \mathbf{W}=\mathbf{1}^{\top}$, and $\rho(\mathbf{W}-\mathbf{1}\mathbf{1}^{\top}/N)<1$, where $\rho(\cdot)$ denotes the spectral radius. 
\end{assumption}
Under Assumption \ref{assump.1}, it is well understood that the iteration in \eqref{eq:distributed-iteration} achieves asymptotic average consensus, i.e., $x_i[t] \to x^* = \frac{1}{N}\sum_{i=1}^{N} x_i[0]$ \cite{200465Lin}.
Moreover, since $\mathbf{W}$ is symmetric, its eigenvalues are real.
Let $\lambda_1, \ldots,\lambda_N$ denote the eigenvalues of $\mathbf{W}$. It follows that $\lambda_1=1$ is a simple eigenvalue, and its associated eigenvector lies in the consensus subspace $\text{span}\{\mathbf{1}\}$, whereas all remaining eigenvalues lie strictly within the interval $(-1,1)$.

\subsection{Assumptions on Link-Level Disturbances}

When communicating over real networks, the information received at each agent might be corrupted by many random factors, such as channel noise, packet losses, or quantization errors. Let $e_i[t] \!\in\! \mathbb{R}$ denote the disturbance observed by agent $i$ at iteration $t$, and $\mathbf{e}[t] \!=\! [e_1[t], \ldots, e_N[t]]^\top$ collects all agent-wise disturbances in a vector. 
Substituting \eqref{noisy_transmitted_signal} into \eqref{noisy_LTI} yields the aggregate disturbance observed at agent $i$ as
\begin{equation}
    e_{i}[t+1] = \sum\nolimits_{j \in \mathcal{N}_i} w_{ij} \epsilon_{ij}[t], \quad\forall t \in \mathbb{N}_0.
\end{equation}

Let $\mathcal{F}_t$ denote the sigma-algebra generated by all information available up to iteration $t \geq 0$. We introduce the following definitions.

\begin{definition}
	\label{definition.Martingale}
	Let $\{\Omega, \, \mathcal{F}, \, \mathbb{P}\}$ be a probability space and $\underline{\mathcal{F}}=\left\{\mathcal{F}_t\right\}_{t \geq 0}$ be a filtration. A stochastic process $\left\{S_t\right\}_{t \geq 0}$ is a martingale with respect to (w.r.t.) $\underline{\mathcal{F}}$ if:
	\begin{itemize}
		\item $S_t$ is $\mathcal{F}_t$-measurable for all $t\geq 0$.
		\item $\mathbb{E}\left[\left|S_t\right|\right]<\infty$ for all $t\geq 0$.
		\item $\mathbb{E}\left[S_{t+1} \mid \mathcal{F}_t\right]=S_t$ for all $t\geq 0$.
	\end{itemize}
\end{definition}

\begin{definition}
	\label{definition.MDS}
	Let $\{\Omega, \, \mathcal{F}, \, \mathbb{P}\}$ be a probability space and $\underline{\mathcal{F}}= \{\mathcal{F}_t\}_{t\geq 0}$ be a filtration. A sequence $\{D[t]\}_{t\geq 1}$ is called a martingale difference sequence (MDS) w.r.t. $\underline{\mathcal{F}}$ if:
	\begin{itemize}
		\item $D[t]$ is $\mathcal{F}_t$-measurable for all $t\geq 1$;
		\item $\mathbb{E}[|D[t]|] < \infty$ for all $t\geq 1$;
		\item $\mathbb{E}[D[t]\mid \mathcal{F}_{t-1}] = 0$ for all $t\geq 1$.
	\end{itemize}
\end{definition}

With the aforementioned definitions, we model the aggregate disturbance $\mathbf e[t]$ according to the following assumption.
\begin{assumption}
	\label{assump.disturbance}
	Assume $\mathbf e[0]=\mathbf 0$ always holds. The sequence $\{\mathbf e[t]\}_{t\geq1}$ is an MDS w.r.t. the filtration $\{\mathcal{F}_t\}_{t\geq0}$ and has uniformly bounded conditional
	second moments, i.e.,
	\begin{equation}
	\sup_{t} \mathbb{E} \left[
	\|\mathbf e[t]\|^2 \mid \mathcal{F}_{t-1}
	\right] \leq \sigma^2 < \infty, \quad \text{almost surely}.
	\end{equation} 
\end{assumption}
Assumption \ref{assump.disturbance} implies that $\mathbf e[t]$ is conditionally zero-mean given the past and its energy is uniformly bounded over time. Note that this model of $\mathbf e[t]$ allows for temporal dependence, time-varying variance, and correlation among agent-wise disturbances, making it more general than  independent and identically distributed (i.i.d.) noise. Note that additive white Gaussian noise (AWGN) is a special case of $\mathbf e[t]$ that satisfies the aforementioned conditions.

\subsection{Average Consensus under Link-Level Disturbances}

The goal of distributed averaging algorithms is for all agents to asymptotically agree on $ x^*=\frac{1}{N}\sum_{i=1}^{N}x_i[0]$.
To characterize the asymptotic behavior of the system, we focus on almost sure convergence, defined as follows.

\begin{definition}
\label{definition.ASAC}[\textbf{Almost sure average consensus {(Def.~4 in \cite{Huang06067359X})}}]
Let $u$ be a random variable with $\mathbb{E}[u] = x^* = \frac{1}{N} \sum_{i=1}^{N}x_i[0]$ and $\operatorname{Var}(u) < \infty$.
The agents are said to achieve \emph{almost sure average consensus} (ASAC), or equivalently, \emph{strong average consensus} if
	\begin{equation}
		\lim _{t \to \infty} x_i[t]=u \quad \text{almost surely}, \quad \forall i\in \mathcal{N}.
	\end{equation}
\end{definition}
A special case arises when $u$ is a constant equal to $x^*$, i.e., $u=\mathbb{E}[u]=x^*$ and $\operatorname{Var}(u)=0$. In this case, the agents are said to achieve \textbf{exact ASAC (E-ASAC)}. In contrast, Definition~\ref{definition.ASAC} only requires the limiting random variable to be unbiased with finite  variance.

Now we briefly recap existing results on average consensus under link-level disturbances using SA techniques. By substituting $e_i[t+1]$ into \eqref{SA}, the SA-based average consensus algorithm can be written as
\begin{align}
	\label{SA1}
	x_i[t +1] 
    &= x_i[t] \!+\! \alpha[t] \sum_{j \in \mathcal{N}_i} w_{ij}\left(y_{ij}[t] \!-\! x_i[t]\right) \\
    &= x_i[t] \!+\! \alpha[t]  \sum_{j \in \mathcal{N}_i} w_{ij} (x_j[t] \!-\! x_i[t]) + \alpha[t] e_{i}[t\!+\!1]. \nonumber 
\end{align}
We can rewrite \eqref{SA1} in matrix form as
\begin{equation}
	\label{Eq.SA_matrix}
	\mathbf{x}[t +1]  = \left(\mathbf{I} - \alpha[t]\mathbf{L}\right) \mathbf{x}[t] + \alpha[t] \, \mathbf{e}[t \!+\!1],
\end{equation}
where $\mathbf{L}$ is a weighted Laplacian matrix with
\begin{equation}
    \label{L_def}
	L_{i j} = 
	\begin{cases}
		\sum_{k \in \mathcal{N}_i} w_{i k}, & i=j, \\ -w_{i j}, & i \neq j \text{ and }(i, j) \in \mathcal{E}, \\ 0, & \text {otherwise} .
	\end{cases}
\end{equation}
By construction, we have $\mathbf L=\mathbf I-\mathbf W$, $\mathbf 1^\top \mathbf L =\mathbf 0$, and  $\mathbf L \mathbf 1=\mathbf 0$.
From \eqref{Eq.SA_matrix}, we define the scaled mixing matrix as 
\begin{equation}
    \mathbf{\tilde{W}}[t]=\mathbf{I}-\alpha[t]\mathbf{L}.
\end{equation}
Clearly, the eigenvalues of $\mathbf{\tilde{W}}[t]$ satisfy 
\begin{equation}
    \lambda_n(\mathbf{\tilde{W}}[t]) = 1-\alpha[t]+\alpha[t]\,\lambda_n, \quad \forall n\in \mathcal{N},
\end{equation} 
where $\{\lambda_n\}_{n=1}^N$ are the eigenvalues of $\mathbf{W}$.
Since $0<\alpha[t]<1$ and $\lambda_n\in(-1,1)$ for all $n\geq2$, it follows that 
\begin{equation}
	\label{contract on disagreement}
	\rho\left(\mathbf{\tilde{W}}[t]\!-\!\frac{\mathbf{1}\mathbf{1}^{\top}}{N}\right) 
	\!=\!\max_{2 \leq n\leq N } |1-\alpha[t]+\alpha[t] \lambda_n| <1, \ \forall t \in \mathbb{N}_0.
\end{equation}
Thus, all eigenvalues of $\widetilde{\mathbf{W}}[t]$ associated with the disagreement subspace $\operatorname{span}\{\mathbf{1}\}^{\perp}$ lie strictly within $(-1,1)$.
Under Assumptions \ref{Assumption: stepsize} and \ref{assump.disturbance} on $\{\alpha[t]\}$ and $\{\mathbf{e}[t]\}$, \eqref{Eq.SA_matrix} achieves ASAC, but generally not E-ASAC \cite{LI5411807}.

\section{DISTRIBUTED OPTIMIZATION FORMULATION OF AVERAGE CONSENSUS}
\label{section.3}

The average consensus problem can be equivalently reformulated as the following distributed optimization problem:
\begin{subequations}\label{Eq.DGD_problem}
\vspace{-3.5mm}
\begin{align}
	\min_{x_1,\ldots,x_N} \  &  \sum\limits_{i=1}^{N} \frac{1}{2}\left(x_i \!-\! x_i[0]\right)^2, \label{Eq.DGD_problem_a} \\
	\text { s.t. } 			\  & x_1=\ldots=x_N. \label{Eq.DGD_problem_b}
    \vspace{-2mm}
\end{align}
\end{subequations}
We define 
\begin{equation}
    f_i(x_i)=\frac{1}{2}\left(x_i \!-\! x_i[0]\right)^2
     \label{eq:local-objective}
\end{equation} as the local objective  of  agent $i$, and $F(x_1,\ldots,x_N)=\sum\nolimits_{i=1}^{N} f_i(x_i)$ as the global objective. Let $\mathbf{x}=[x_1,\ldots,x_N]^\top$ be the optimization variables in a vector form. Under the consensus constraint, the unique minimizer of the global objective function is $\mathbf{x}^*=x^{*}\mathbf{1}$, where $x^{*}=\frac{1}{N} \sum_{i=1}^{N}x_i[0]$.

One of the standard methods for distributed optimization is decentralized gradient descent (DGD) \cite{4749425Nedic}, which combines consensus iterations with local gradient updates. Under noisy communication, applying SA with diminishing stepsizes to the consensus iterations can mitigate the impact of link disturbances \cite{Touri5203882}. Combining these ideas leads to a two-stepsize scheme for noisy distributed optimization \cite{Srivastava5717947}:
\begin{align}
	\label{Eq.DGD}
	\mathbf{x}[t \! + \! 1] &= \left(\mathbf{I} - \alpha[t]\mathbf{L}\right)\! \mathbf{x}[t] - \eta[t] \nabla F(\mathbf{x}[t]) + \alpha[t] \mathbf{e}[t \!+\!1] \nonumber \\
    &= \mathbf{\tilde{W}}[t] \mathbf{x}[t] - \eta[t] \nabla F(\mathbf{x}[t]) + \alpha[t] \mathbf{e}[t \!+\!1],
\end{align}
where $\nabla F(\mathbf{x}[t]) = [\nabla f_{1}(x_1[t]),\, \ldots,\, \nabla f_{N}(x_N[t])]^{\top}$ is the gradient of local objective functions stacked in a vector. 
The gradient stepsize $\eta[t]$ is assumed to satisfy the same conditions as $\alpha[t]$ in Assumption \ref{Assumption: stepsize}.
Given the local objectives defined in \eqref{eq:local-objective}, we have
\begin{equation}
	\label{Eq.local_gradient_matrix}
	\nabla F(\mathbf{x}[t]) = \mathbf{x}[t] - \mathbf{x}[0].
\end{equation}
Substituting \eqref{Eq.local_gradient_matrix} into \eqref{Eq.DGD}, we obtain
\begin{align}
	\label{Eq.Anchoring_AC}
	\mathbf{x}[t \! + \! 1] 
	&= \mathbf{\tilde{W}}[t] \mathbf{x}[t] \!-\! \eta[t] \!\left(\mathbf{x}[t] \!-\! \mathbf{x}[0]\right) \!+\! \alpha[t]  \mathbf{e}[t \!+\!1] \\
	&= \left((1\!-\!\eta[t])\mathbf{I} \!-\! \alpha[t]\mathbf{L}\right)  \mathbf{x}[t] \!+\! \eta[t] \mathbf{x}[0] \!+\! \alpha[t]  \mathbf{e}[t \!+\!1]. \nonumber
\end{align}
The resulting update rule can be interpreted as augmenting the SA-based iteration in \eqref{Eq.SA_matrix} with an \textit{anchoring} term $-\eta[t](\mathbf{x}[t]-\mathbf{x}[0])$, which continuously pulls each agent state toward its initial value. We refer to \eqref{Eq.Anchoring_AC} as the \textbf{anchor-based average consensus (AAC)} algorithm. The classical SA-based iteration in \eqref{Eq.SA_matrix} is termed the \textbf{primal average consensus (PAC)} algorithm in the remainder of this paper. Additionally, \eqref{Eq.Anchoring_AC} can also be interpreted as a leakage-based average consensus algorithm equipped with diminishing stepsizes $\{\alpha[t]\}$ and $\{\eta[t]\}$, or equivalently, as a combination of the SA-based and leakage-based mechanisms in \cite{Pescosolido4641610}.

Define $\mathbf A_t \!\triangleq\!\! (1\!-\!\eta[t])\mathbf{I} \!-\! \alpha[t]\mathbf{L}$. To ensure asymptotic contraction of the disagreement component, we require that the spectral radius satisfies\footnote{Generally, it is not necessary to require \eqref{contract on disagreement_AAC} to hold for all $t \in \mathbb{N}_0$. Since $\alpha[t] \to 0$ and $\eta[t]\to0$, there exists a finite $T$ such that \eqref{contract on disagreement_AAC} is satisfied for all $t > T$. For asymptotic analysis, this tail condition is sufficient.}:
\begin{equation}
	\label{contract on disagreement_AAC}
	\rho\left(\!\mathbf{A}_t \!-\!\frac{\mathbf{1}\mathbf{1}^{\top}}{N}\!\right) \!=\!\!\! \max_{2 \leq n\leq N } \! |1\!-\!\eta[t]\!-\!\alpha[t]\!+\!\alpha[t] \lambda_n| \!<\! 1, \, \forall t \in \mathbb{N}_0.
\end{equation}
Although \eqref{contract on disagreement_AAC} guarantees that \(\mathbf{A}_t\) is contractive on the disagreement subspace, the anchoring term introduces an additional effect of order $\eta[t] $ that keeps pulling each agent toward its initial state. To ensure that the anchoring effect is asymptotically negligible compared with the contraction in the disagreement subspace, we further require
\begin{equation}
	\label{stepsizes relation}
	\sum\nolimits_{t=0}^{\infty}\eta[t]^2/\alpha[t]<\infty.
\end{equation}

\begin{remark}
From a distributed optimization perspective, \eqref{Eq.Anchoring_AC} is a  two-stepsize noisy DGD-type recursion \cite{Srivastava5717947} applied to the average-consensus problem defined in \eqref{Eq.DGD_problem}. In this problem, the local gradient correction $-\eta[t](\mathbf{x}[t]-\mathbf{x}[0])$ acts as an anchor toward the initial states. 
Since \eqref{Eq.DGD_problem} admits a unique optimizer $\mathbf{x}^*$, this anchor steers the iterates toward the exact initial average. Under the diminishing stepsizes and the MDS disturbance assumptions, AAC achieves E-ASAC. In contrast, PAC lacks such a mechanism, so persistent disturbances accumulate over iterations, resulting in consensus at a random value with nonzero variance.
\end{remark}

\section{STATE ANCHORING: FROM DRIFT ANALYSIS TO UNIFIED INTERPRETATION}

Although \eqref{Eq.Anchoring_AC}  is structurally related to existing noisy DGD methods with decaying stepsizes, we provide a self-contained convergence analysis here by studying its linear iteration form.
To start with, we introduce a projection $\mathbf J$ onto the  consensus subspace, and an orthogonal projector $\mathbf P$ onto the disagreement subspace:
\begin{equation}
	\label{subspace decomposition}
	\mathbf J \triangleq \frac{1}{N}\mathbf{1}\mathbf{1}^{\top}, \quad \mathbf P \triangleq \mathbf I - \mathbf J = \mathbf I - \frac{1}{N}\mathbf{1}\mathbf{1}^{\top}.
\end{equation}
Moreover, we define the network average at iteration $t$ as
\begin{equation}
	\label{network average} 
	m[t] \triangleq \frac{1}{N}\mathbf{1}^{\top} \mathbf x[t].
\end{equation}
Immediately, we have:
\begin{equation}
	\label{average subspace} 
	\mathbf J \, \mathbf x[t] = m[t] \, \mathbf 1 \in \mathbb R^{N}.
\end{equation}
Similarly, we define the network disagreement as
\begin{equation}
	\label{network disagreement} 
	\mathbf d[t] \triangleq \mathbf P \, \mathbf x[t] \in \mathbb R^{N}.
\end{equation}
Thus every state value vector admits a decomposition 
\begin{equation}
    \mathbf x[t] \triangleq m[t]\, \mathbf{1} + \mathbf d[t], \quad \text{with}\  \mathbf d[t] \in \operatorname{span}\{\mathbf 1\}^{\perp}.    
\end{equation} 
It follows that the squared Euclidean distance between $\mathbf x[t]$ and the network initial average $\mathbf{x}^* \!=\! m[0] \mathbf 1$ admits an orthogonal decomposition:
\begin{align}
	\label{Decomposition}
	&\left\| \mathbf x[t] - \mathbf x^* \right\|^2 \nonumber \\ 
    =&  \left\| \mathbf J \mathbf x[t] - \mathbf x^* + \mathbf P \mathbf x[t]\right\|^2  = \|\left(m[t] - m[0]\right)\mathbf{1} + \mathbf d[t] \|^2 \nonumber \\
    =& \underbrace{N\left(m[t] - m[0]\right)^2}_{\textbf{average bias}} + \underbrace{\|\mathbf d[t]\|^2}_{\textbf{disagreement}}.
\end{align}

The asymptotic behaviors of AAC and PAC are different: AAC achieves E-ASAC, while PAC can only achieve ASAC. These two results are stated formally in Theorem \ref{theorem.AAC&PAC}, and their proofs are provided in the next two subsections.
\begin{theorem}
\label{theorem.AAC&PAC}[\textbf{Convergence of AAC and PAC}]
Suppose that Assumptions \ref{Assumption: stepsize}--\ref{assump.disturbance} hold, $\{\eta[t]\}$ satisfies the
same conditions as $\{\alpha[t]\}$ in Assumption \ref{Assumption: stepsize}, and \eqref{stepsizes relation} holds.
Then the AAC algorithm \eqref{Eq.Anchoring_AC} achieves E-ASAC: 
\begin{equation}
    \lim _{t \to \infty} \left\| \mathbf x[t] - \mathbf x^* \right\|^2=0 \ \Rightarrow\  \lim _{t \to \infty}  \mathbf x[t]= \mathbf x^*\  \text{almost surely.} \nonumber
\end{equation}
The PAC algorithm \eqref{Eq.SA_matrix} achieves ASAC to a random variable $u \triangleq \lim_{t \to \infty}m[t] \mathbf\, \in \mathbb{R}$ satisfying
\begin{equation}
    \mathbb E[u] = x^*,  \quad 0<\operatorname{Var}(u) < \infty. \footnote{The strict positivity of $\operatorname{Var}(u)$ requires the disturbance to have a non-degenerate projection onto the consensus subspace, i.e., $\bar e[t]=\frac{1} {N}\mathbf{1}^{\top}\mathbf e[t]$ is non-degenerate for all $t\geq 1$.} \nonumber
\end{equation}
That is, 
\begin{equation}
    \lim _{t \rightarrow \infty} \left\| \mathbf x[t] - u\mathbf 1 \right\|^2 = 0 \ \Rightarrow\  \lim _{t \to \infty}  \mathbf x[t]= u \mathbf 1\  \text{almost surely.} \nonumber
\end{equation}
\end{theorem}

\subsection{Random Drift without Anchoring}

We first recall the following classical almost-sure convergence result for stochastic recursions, which will be used in the analysis of both PAC and AAC.
\begin{lemma}
	\label{lemma:RS_lemma}[\textbf{Robbins--Siegmund Corollary (Lemma 5.31 in \cite{BauschkeCombettes2017})}]
	Let $\{\Omega, \, \mathcal{F}, \, \mathbb{P}\}$ be a probability space and $\underline{\mathcal{F}}=\left\{\mathcal{F}_t\right\}_{t \geq 0}$ be a filtration. 
	Let $\{a_t\}$ be a sequence of nonnegative integrable random variables adapted to $\underline{\mathcal{F}}$, and $\{b_t\}$ and $\{c_t\}$ be nonnegative deterministic sequences such that $\sum_{t=0}^{\infty}b_t = \infty$ and $\sum_{t=0}^{\infty}c_t < \infty$. Suppose that 
	\begin{equation}
		\mathbb{E}[a_{t+1}\mid \mathcal{F}_t] \le (1-b_t)a_t + c_t, \quad \forall t\in \mathbb{N}_0,
	\end{equation}
	almost surely. Then
	\begin{equation}
		\lim_{t\to\infty}	a_t = 0 \quad \text{almost surely}.
	\end{equation}
\end{lemma}
We now analyze the asymptotic behavior of $\|\mathbf d[t]\|^2$ for PAC. Applying the projector $\mathbf P$ to both sides of \eqref{Eq.SA_matrix} yields
\begin{equation}
	\label{PAC-disagreement}
	\mathbf{d}[t \!+\!1] = \mathbf{P} \left(\mathbf{I} - \alpha[t]\mathbf{L}\right) \mathbf x[t] + \alpha[t] \mathbf P\mathbf e[t \!+\!1].
\end{equation}
Since $\mathbf L\mathbf 1=\mathbf 0$, $\mathbf L$ is symmetric, and $\mathbf{\tilde{W}}[t]=\mathbf{I}-\alpha[t]\mathbf{L}$, we have $\mathbf P\mathbf{\tilde{W}}[t]=\mathbf{\tilde{W}}[t]\mathbf P$ and hence \eqref{PAC-disagreement} becomes
\begin{equation}
	\label{PAC-disagreement2}
	\mathbf{d}[t \!+\!1] = \mathbf{\tilde{W}}[t]\mathbf d[t]+\alpha[t]\mathbf P\mathbf e[t \!+\!1].
\end{equation}

Given that $\mathbf d[t] \in \operatorname{span}\{\mathbf 1\}^{\perp}$ and $\mathbf{\tilde{W}}[t]$ is strictly contractive on the disagreement subspace as shown in \eqref{contract on disagreement}, the term $\mathbf{\tilde{W}}[t]\mathbf d[t]$ in \eqref{PAC-disagreement2} is asymptotically attenuated. 
Specifically, define $\beta[t] \coloneq \max_{2 \leq n\leq N }|1-\alpha[t]+\alpha[t] \lambda_n| \in [0,1)$, then  
\begin{equation}
	\left\|\mathbf{\tilde{W}}[t]\, \mathbf{d}[t]\right\| \leq \beta[t]\|\mathbf{d}[t]\|.
\end{equation}

Since $\|\mathbf{P} \mathbf{e}[t \!+\!1]\| \leq \|\mathbf{e}[t \!+\!1]\|$, taking conditional expectation of the squared norm on both sides of $\eqref{PAC-disagreement2}$ yields
\begin{align}
	\label{PAC-disagreement3}
	\mathbb{E} [\|\mathbf{d}[t \!+\!1]\|^2 \!\mid\! \mathcal{F}_t] \!
    &\leq \beta[t]^2 \|\mathbf{d}[t]\|^2 \!+\! \alpha[t]^2  \mathbb{E}[\|\mathbf{e}[t \!+\!1]\|^2 \!\mid\! \mathcal{F}_t] \nonumber \\
    & \leq (1-c\alpha[t]) \|\mathbf{d}[t]\|^2+\alpha[t]^2\, \sigma^2,
\end{align}
where $c>0$ is a small and positive constant satisfying $\beta[t]^2 \leq 1-c\,\alpha[t]$. The last inequality of \eqref{PAC-disagreement3} follows from the assumption that $\sup _t \mathbb{E}[\left\|\mathbf e[t]\right\|^2 \!\mid\! \mathcal{F}_{t-1}] \leq \sigma^2$. 
By Assumption \ref{Assumption: stepsize}, applying Lemma \ref{lemma:RS_lemma} to \eqref{PAC-disagreement3} yields
\begin{equation}
	\label{PAC-disagreement5}
    \lim_{t\to\infty}\|\mathbf d[t]\|^2 =0 \quad \text{almost surely}.
\end{equation}

Next, we study the average bias term in \eqref{Decomposition} under the PAC algorithm. As will be shown, its network average is driven by persistent disturbances and exhibits a random-walk type behavior. This prevents PAC from achieving E-ASAC, even if the disagreement component vanishes asymptotically.

To see this, left-multiplying by $\frac{1}{N} \mathbf{1}^{\top}$ on both sides of \eqref{Eq.SA_matrix} and using $\mathbf{1}^{\top} \mathbf{\tilde{W}}[t] = \mathbf{1}^{\top}$, we obtain:
\begin{equation}
	\label{PAC_average0}
	m[t \!+\! 1] = m[t] + \frac{\alpha[t]}{N}\, \mathbf 1^{\top}\, \mathbf e[t \!+\! 1].
\end{equation}
Let $\bar{e}[t]$ denote the agent-wise average of $\mathbf e[t]$:
\begin{equation}
	\label{Error_average}
	 \bar{e}[t] \triangleq \frac{1}{N} \mathbf 1^{\top}\, \mathbf e[t],
\end{equation}
so that \eqref{PAC_average0} can be re-written as:
\begin{equation}
	\label{PAC_average1}
	m[t \!+\! 1] = m[t] + \alpha[t] \, \bar{e}[t \!+\! 1].
\end{equation}
Since $\bar{e}[t]$ is a linear transformation of $\mathbf{e}[t]$, the sequence $\{\bar{e}[t]\}$ is also a MDS. Consequently, \eqref{PAC_average1} defines a martingale recursion where $\{m[t]\}$ is a martingale w.r.t. $\{\mathcal{F}_t\}$. Given that $\sup _t \mathbb{E}[\left\|\mathbf e[t]\right\|^2 \!\mid\! \mathcal{F}_{t-1}] \leq \sigma^2$,  
\begin{equation} 
	\label{average error bound}
	\mathbb{E}[(\bar{e}[t])^2  \mid \mathcal{F}_{t-1} ] \leq \frac{1}{N} \mathbb{E}[\left\|\mathbf e[t]\right\|^2 \mid \mathcal{F}_{t-1}] \leq \sigma^2/N,
\end{equation}
holds. Furthermore, \eqref{PAC_average1} can be rewritten as
\begin{equation}
	\label{PAC_average2}
	m[t] = m[0] + \sum\nolimits_{k=1}^{t} \alpha[k\!-\!1] \, \bar{e}[k].
\end{equation}

Let $S[t] \!\coloneq\! \sum_{k=1}^{t} \alpha[k\!-\!1]\bar{e}[k]$, representing the network average bias in \eqref{Decomposition}. Since $\{\bar e[k]\}$ is a MDS and  $\alpha[k\!-\!1]$ is deterministic, each increment $\alpha[k\!-\!1]\bar e[k]$ is $\mathcal{F}_{k}$-measurable and satisfies
\begin{equation}
	\label{ST_increment}
	\mathbb{E} \left[\alpha[k\!-\!1] \bar e[k] \!\mid\! \mathcal{F}_{k-1}\right] = \alpha[k\!-\!1]\mathbb{E}\left[\bar e[k] \!\mid\! \mathcal{F}_{k-1}\right] =0.
\end{equation}
Hence, $S[t]$ is $\mathcal{F}_t$-measurable and
\begin{equation}
	\mathbb{E}[S[t] \!\mid\! \mathcal{F}_{t -\!1}]
	\!=\!
	\mathbb{E}\left[S[t \!-\!1] \!+\! \alpha[t \!-\! 1]\bar e[t] \!\mid\! \mathcal{F}_{t -\!1}\right]
	\!=\! S[t \!-\! 1].
\end{equation}
Moreover, we have $S[0] = \bar{e}[0] = 0$. Given the conditional zero-mean increments \eqref{ST_increment}, $\mathbb{E}[S[t]]=0$ holds for all $t$. Therefore, $\{S[t]\}$ is a zero-mean martingale w.r.t. $\{\mathcal{F}_t\}$. Taking the expectation of $S[t]^2$, using the upper bound \eqref{average error bound}, and the orthogonality of martingale increments, we get
\begin{align}
	\label{con_exp S[t]}
	\mathbb{E}[S^2[t]] &= \mathbb{E}\left[\mathbb{E}[S[t]^2 \mid \mathcal{F}_{t-1}]\right] \nonumber \\
    &= \mathbb{E}[S[t\!-\!1]^2] + \mathbb{E}\left[\alpha[t\!-\!1]^2 \mathbb{E}\left[(\bar{e}[t])^2 \mid \mathcal{F}_{t-1}\right] \right] \nonumber \\
    & \quad+ \mathbb{E}\left[2\alpha[t\!-\!1]S[t\!-\!1]\mathbb{E}\left[\bar{e}[t] \mid \mathcal{F}_{t-1}\right] \right] \nonumber \\
    &\leq \mathbb{E}[S[t-1]^2]+\frac{\alpha[t\!-\!1]^2\sigma^2}{N} \nonumber \\
    &\leq \frac{\sigma^2}{N} \sum_{k=1}^{t}\alpha[k- 1]^2.
\end{align}

Under the non-degeneracy condition of $\bar e[t]$ stated in Theorem \ref{theorem.AAC&PAC}, we have $\mathbb{E}[S[t]^2] > 0$ for all $t>0$.
Since $\sum_{k=0}^{\infty}\alpha[k]^2 < \infty$, it follows that $\sup_t \mathbb{E}[S[t]^2] <\infty$. This implies that $\{S[t]\}$ is an $L^2$-bounded martingale, thereby converging almost surely and in mean square to a random variable $S[\infty]$. Therefore, $m[t]$ converges almost surely, i.e.,
\begin{equation}
    \label{average_limit}
	\lim_{t \to \infty}m[t] = m[0] + S[\infty] =  m[\infty]  \quad  \text{almost surely}.
\end{equation}

Since $\{S[t]\}$ is an $L^2$-bounded martingale with zero-mean increments, one readily obtains
\begin{align}
    \label{average_expectation}
	& \mathbb{E}[m[\infty]] = m[0] + \mathbb{E}[S[\infty]] = m[0]; \\  	
    \label{average_variance}
	& \operatorname{Var}(m[\infty]) = \mathbb{E}[S[\infty]^2] \in \left(0,\,\infty\right). 
\end{align}

Consequently, the network average under PAC converges to a random limit $u \coloneq m[\infty]$ centered at the initial network average $m[0]$, but not to $m[0]$ itself. 
By \eqref{PAC-disagreement5} and \eqref{average_limit}, we have $m[t] \!\to\! u$ and $\|\mathbf d[t]\|^2 \!\to\! 0$ almost surely. Applying the decomposition \eqref{Decomposition} yields
\begin{equation}
    \label{PAC_result}
    \lim _{t \rightarrow \infty} \left\| \mathbf x[t] \!-\! u\mathbf 1 \right\|^2 \!=\! N\left(m[t] \!-\! u\right)^2 \!+\! \|\mathbf d[t]\|^2 \!=\! 0
\end{equation}
almost surely. It follows that
\begin{equation}
	\label{PAC_result2}
	\mathbf x[t] \to  u \mathbf1  \quad \text{almost surely.}
\end{equation}

Therefore, combining \eqref{PAC-disagreement5} and \eqref{average_limit}--\eqref{PAC_result}, we conclude that the PAC algorithm achieves ASAC w.r.t. the random variable $u$, which is in line with the result of the SA-based average consensus algorithm in \cite{LI5411807}. Our decomposition further shows that the accumulation of disturbances induces a random drift in the network average, which leads to the residual consensus error.

\subsection{The Effects of Anchoring: How Does It Help?}
The failure of PAC to achieve E-ASAC stems from the fact that the average disturbance $\bar{e}[t]$ enters the network average $m[t]$  and accumulates over time, thereby resulting in a random drift. To address this effect, AAC introduces a contraction factor $1-\eta[t]$ on the average bias $m[t]-m[0]$, hence driving it back to $0$. To see this, we define 
\begin{equation}
    \Delta[t] \coloneq m[t]-m[0].
\end{equation}
Left-multiplying both sides of \eqref{Eq.Anchoring_AC} by $\frac{1}{N}\mathbf{1}^\top$ and using the property $\mathbf{1}^\top \mathbf{\tilde{W}}[t]=\mathbf{1}^\top$, we obtain
\begin{equation}
	\label{Eq.average_bias}
	\Delta[t+1]=(1-\eta[t])\Delta[t]+\alpha[t]\bar e[t+1].
\end{equation}
Since $\{\eta[t]\}$ satisfies $\sum_{t=0}^{\infty} \eta[t]=\infty$ and $0\!<\!\eta[t]\!<\!1$,
\begin{equation}
\label{contraction}
\prod_{t=0}^{\infty}(1-\eta[t]) = 0.
\end{equation}
Note that \eqref{contraction} holds from the inequality $\log(1-\eta[t]) \leq -\eta[t]$ for all $t$, which implies
$$\log\big(\prod_{t=0}^{\infty}(1\!-\!\eta[t])\big) \!=\! \sum_{t=0}^{\infty}\log(1\!-\!\eta[t]) \!\leq\! -\!\sum_{t=0}^{\infty}\eta[t] \!=\! -\infty.$$
Therefore, the anchoring term induces a persistent contraction that can offset the accumulated martingale disturbance. This observation captures the key mechanism by which AAC drives $m[t]\to m[0]$ almost surely. 

To establish this claim rigorously, we take conditional expectation of the squared value on both sides of \eqref{Eq.average_bias}:
\begin{align}
	\label{con_exp_ave_bias}
	&\mathbb E\left[\Delta[t+1]^2\mid \mathcal F_t\right] \nonumber \\
	=& (1-\eta[t])^2 \Delta[t]^2 + \alpha[t]^2\,\mathbb E\left[(\bar e[t+1])^2\mid \mathcal F_t\right]. \nonumber \\
    \leq&  (1-\eta[t]) \Delta[t]^2 +\frac{\alpha[t]^2 \sigma^2}{N},
\end{align}
where the last inequality holds from \eqref{average error bound}. By Assumption \ref{Assumption: stepsize} on both $\{\alpha[t]\}$ and $\{\eta[t]\}$, applying Lemma \ref{lemma:RS_lemma} to \eqref{con_exp_ave_bias} yields,
\begin{equation}
	\label{AAC_average_bias}
    \lim_{t\to\infty}\Delta[t]^2 = \lim_{t\to\infty} (m[t]-m[0])^2 = 0\quad \text{almost surely}.
\end{equation}
This implies that the average bias term under AAC converges to zero almost surely.

Next, to complete the convergence proof, we analyze the disagreement term. By applying the projector $\mathbf{P}$ to both sides of \eqref{Eq.Anchoring_AC}, one readily obtains 
\begin{align}
	\label{AAC-disagreement2}
	\mathbf{d}[t \!+\!1] 
	&= \left(\mathbf{I} \!-\! \alpha[t]\mathbf{L}\right)\mathbf d[t] - \eta[t] \!\left(\mathbf d[t]\! -\! \mathbf d[0]\right) +\alpha[t]\mathbf P\mathbf e[t \!+\!1] \nonumber \\
	&= \mathbf{A}_t \mathbf d[t] + \eta[t] \mathbf d[0] +\alpha[t]\mathbf P\mathbf e[t \!+\!1].
\end{align}
Since $\|\mathbf{P} \mathbf{e}[t \!+\!1]\| \leq \|\mathbf{e}[t \!+\!1]\|$ and $\sup _t \mathbb{E}[\left\|\mathbf e[t]\right\|^2 \mid \mathcal{F}_{t-1}] \leq \sigma^2$, taking the conditional expectation of the squared norm on both sides of $\eqref{AAC-disagreement2}$ yields
\begin{align}
	\label{AAC-disagreement3}
	&\mathbb{E}[ \| \mathbf{d}[t \!+\!1]\|^2\!\mid\! \mathcal{F}_t] \nonumber \\ 
	=& \|\mathbf{A}_t \mathbf d[t] + \eta[t] \,\mathbf d[0] \|^2 +\alpha[t]^2 \,\mathbb{E}[\|\mathbf{P} \mathbf{e}[t \!+\!1]\|^2 \!\mid\! \mathcal{F}_t]. \nonumber \\ 
    \leq& \|\mathbf{A}_t \mathbf d[t] + \eta[t] \, \mathbf d[0] \|^2
	+\alpha[t]^2 \sigma^2.
\end{align}

We now focus on bounding the first term on the right-hand side of \eqref{AAC-disagreement3}, $\|\mathbf{A}_t \mathbf d[t] \!+\! \eta[t] \mathbf d[0] \|^2$.
By Young’s inequality, for any $\varepsilon_t \! >\! 0$, we have
\begin{align}
	\label{AAC_linear_part}
	&\|\mathbf{A}_t \mathbf d[t] + \eta[t] \,\mathbf d[0] \|^2 \nonumber \\
	\leq& (1+\varepsilon_t)\left\|\mathbf{A}_t \mathbf d[t]\right\|^2+\left(1+\frac{1}{\varepsilon_t}\right) \eta[t]^2\|\mathbf d[0]\|^2. 
\end{align}
By \eqref{contract on disagreement_AAC}, define 
\begin{equation}
    \zeta[t]^2 \coloneq \max_{2 \leq n\leq N }|1\!-\!\eta[t]\!-\!\alpha[t]\!+\!\alpha[t] \lambda_n|^2 \in [0,1),
\end{equation} 
so that $\|\mathbf{A}_t \mathbf d[t]\| \leq \zeta[t]\|\mathbf d[t]\|$. 
For a sufficiently small constant $c_0$, we have $\zeta[t]^2 \leq 1-c_0\, (\eta[t] + \alpha[t])$. 

Now substituting the above bounds into \eqref{AAC_linear_part} with the choice $\varepsilon_t \triangleq \frac{c_0}{2}\, (\eta[t] + \alpha[t])$, we have
\begin{align}
	\label{AAC_linear_part2}
	&\|\mathbf{A}_t \mathbf d[t] + \eta[t] \,\mathbf d[0] \|^2 \nonumber \\
	\leq& \big(1+\frac{c_0}{2}(\eta[t] + \alpha[t])\big) \left(1-c_0 (\eta[t] + \alpha[t])\right) \|\mathbf d[t]\|^2 \nonumber \\ &+\eta[t]^2 \|\mathbf d[0]\|^2 + \frac{2\eta[t]^2}{c_0(\eta[t] + \alpha[t])} \|\mathbf d[0]\|^2  \\
	\leq& \big(1\!-\!\frac{c_0}{2}(\eta[t] \!+\! \alpha[t])\big) \!\|\mathbf d[t]\|^2 \nonumber \\
    &+ \big(\eta[t]^2 \!+\! \frac{2\eta[t]^2}{c_0(\eta[t] + \alpha[t])}\big) \! \|\mathbf d[0]\|^2. \nonumber
\end{align}
The last inequality follows by applying $(1+ax)(1-bx) = 1+(a-b)x-abx^2 \leq 1+(a-b)x$ with $2a = b = c_0$ and $x = (\eta[t] \!+\! \alpha[t])$. Substituting \eqref{AAC_linear_part2} into \eqref{AAC-disagreement2} and using the fact that $\eta[t]^2/(\eta[t] + \alpha[t]) \leq \eta[t]^2/\alpha[t]$ for all $t$, we obtain
\begin{align}
	\label{AAC-disagreement4}
	&\mathbb{E}[ \| \mathbf{d}[t \!+\!1]\|^2 \mid \mathcal{F}_t] \\
	\leq& \big(1-\frac{c_0(\eta[t] + \alpha[t])}{2}\big)  \|\mathbf d[t]\|^2 \nonumber \\
    &+ \eta[t]^2 \|\mathbf d[0]\|^2 + \frac{\eta[t]^2}{\alpha[t]} \frac{2\|\mathbf d[0]\|^2}{c_0} + \alpha[t]^2 \sigma^2. \nonumber
\end{align}
Since 
\begin{align}
&\sum_{t=0}^{\infty}  \frac{c_0}{2}(\eta[t] + \alpha[t])=\infty, \\ &\sum_{t=0}^{\infty}  \eta[t]^2 \|\mathbf d[0]\|^2 + \frac{\eta[t]^2}{\alpha[t]} \frac{2\|\mathbf d[0]\|^2}{c_0} + \alpha[t]^2 \sigma^2 <\infty,
\end{align}
applying Lemma \ref{lemma:RS_lemma} to \eqref{AAC-disagreement4} yields 
\begin{equation}
	\label{AAC-disagreement5}
	\lim_{t\to\infty}\|\mathbf d[t]\|^2 =0 \quad \text{almost surely}.
\end{equation}
Combining \eqref{AAC_average_bias} and \eqref{AAC-disagreement5} with the decomposition in \eqref{Decomposition}, we conclude that 
\begin{align}
	\label{AAC-convergence}
	&\lim _{t \rightarrow \infty} \left\| \mathbf x[t] - \mathbf x^* \right\|^2 \nonumber \\
    =& \lim _{t \rightarrow \infty} \!N\!\left(m[t] \!-\! m[0]\right)^2 \!+\! \|\mathbf d[t]\|^2 \!=\! 0 \ \ \, \text{almost surely. }
\end{align}
It follows that
\begin{equation}
	\label{AAC-convergence2}
	\mathbf x[t] \to \mathbf x^*  \quad \text{almost surely,}
\end{equation}
i.e., the AAC algorithm achieves E-ASAC. The key to the proof is to decompose \eqref{Decomposition} into an average-bias term and a disagreement term: \eqref{AAC_average_bias} shows that the average bias vanishes almost surely, whereas \eqref{AAC-disagreement5} establishes the same result for the disagreement. Intuitively, the consensus update contracts the disagreement subspace, while the anchoring term prevents drift from the initial network average.

\subsection{Unified Interpretation of Representative Methods}

The proposed AAC algorithm provides a unified interpretation for understanding several representative noise-robust consensus schemes, including noise-resilient push-sum (NR-PushSum) \cite{Vivek10286415}, robust dynamic average consensus \cite{Wang10383541}, and non-coherent over-the-air (OTA) distributed consensus \cite{Deng11161728}.

From the anchoring perspective, these methods are not isolated algorithmic constructions, but different realizations of a common principle: \textbf{introducing an anchor toward the desired target to suppress noise accumulation}.

NR-PushSum \cite{Vivek10286415} modifies the classical push-sum recursion by repeatedly injecting weighted initial values into the updates, which indeed serve as anchors. Its iteration can be compactly represented as
\begin{equation}
    \label{Eq.NRPushSum}
	\begin{aligned}
		\mathbf{x}[t+1]
		&=\mathbf{W}_{\mathrm{NR}}[t]\mathbf{x}[t]
		+\eta[t]\mathbf{x}[0]+\alpha[t]\mathbf{e}_{x}[t],\\
		\mathbf{y}[t+1]
		&=\mathbf{W}_{\mathrm{NR}}[t]\mathbf{y}[t]
		+\eta[t]\mathbf{y}[0]+\alpha[t]\mathbf{e}_{y}[t],\\
		z_i[t+1]
		&=x_i[t+1]/y_i[t+1],\quad i\in\mathcal{N},
	\end{aligned}
\end{equation}
where $\mathbf{W}_{\mathrm{NR}}[t]$ denotes the mixing matrix of NR-PushSum, $\mathbf{e}_x[t]$ and $\mathbf{e}_y[t]$ are communication disturbances, $\mathbf{x}[t]$ is the state information, $\mathbf{y}[t]$ provides the normalization weights, and $z_i[t]$ is the resulting consensus estimate. Indeed, it can be viewed as the push-sum counterpart of the anchoring mechanism over directed graphs. Moreover, NR-PushSum can also be interpreted as a modified subgradient-push method \cite{Nedic6930814}, where the local subgradient term plays an anchoring role. 

A similar interpretation applies to dynamic average consensus \cite{Wang10383541}. Instead of anchoring the iterates to a time-invariant initial value, it adopts a time-varying reference signal, enabling the network to track an evolving target while remaining robust to noise. Its update can be represented as
\begin{align}
	x_i[t+1] = &x_i[t] + \text{ neighbor interaction } \nonumber \\ 
	& \underbrace{-\eta[t]\big(x_i[t]-r_i[t]\big)}_{\text{anchoring}}
	+
	\underbrace{\big(r_i[t+1]-r_i[t]\big)}_{\text{reference variation}},
	\label{Eq.Dynamic_anchor}
\end{align}
where $r_i[t]$ is the reference signal. The anchoring term has the same structure as the one in \eqref{Eq.Anchoring_AC}, with the fixed $x_i[0]$ replaced by a time-varying $r_i[t]$. Therefore, it can be viewed as a dynamic counterpart of the anchoring mechanism. This observation suggests that anchoring is not tied to a specific algorithmic structure; rather, it is a general mechanism for keeping the system centered around a desired target and suppressing the accumulation of persistent noise.

Our earlier work on average consensus via non-coherent OTA aggregation \cite{Deng11161728} did not explicitly formulate AAC as its main theme, but it implicitly employed the same anchoring idea derived from a decentralized projected gradient descent update. In particular, its iteration is given by
\begin{equation}
    \label{Eq.OTA_anchor}
	\mathbf{x}[t+1]
	=
	\mathcal{P}_{\Omega}\!\left[
	\mathbf{W}[t]\mathbf{x}[t]
	-
	\eta[t]\big(\mathbf{x}[t]-\mathbf{x}[0]\big)
	\right],
\end{equation}
where $\mathcal{P}_{\Omega}[\cdot]$ is a projection onto a feasible set, and $\mathbf{W}[t]$ is induced by OTA aggregation.
As shown in \eqref{Eq.OTA_anchor}, the non-coherent OTA consensus setting has the same anchoring structure as the AAC algorithm \eqref{Eq.Anchoring_AC}.
This result provides an engineering realization of the anchoring mechanism in a noisy wireless setting.

Taken together, these methods support a unified interpretation: anchoring is a key mechanism in robust consensus design under persistent noise. This idea extends naturally to dynamic consensus, directed graphs, and practical wireless communication scenarios.

\section{NUMERICAL SIMULATIONS}
In this section, we illustrate the performance of the AAC and PAC algorithms through numerical experiments. The communication network is modeled as one instance of a connected undirected Erd\H{o}s--R\'enyi graph with $N=36$ agents and  link probability of $0.5$. The initial states are independently drawn from the uniform distribution on $[-100, 100]$. We consider an equal-weight mixing matrix design, i.e.,
\begin{align}
    w_{ij} = \begin{cases}
        \epsilon  \quad &\text{if } (i,j)\in\mathcal{E}, i\neq j,\\
        1-\sum_{j\in\mathcal{N}_i}w_{ij} \quad  &\text{if } i=j,
    \end{cases}
\end{align}
with $\epsilon=\frac{1}{\max\{d_i\}+1}$, where $d_i$ is the degree of agent $i$.

The disturbance sequence $\{\mathbf{e}[t]\}$ is introduced by random quantization noise. Specifically, each agent $i$ transmits a quantized value of the current state $\hat{x}_i[t] = \mathbf Q ( x_i[t])$, where $\mathbf Q(\cdot)$ is a $3$-bit unbiased stochastic quantizer with input range $[-100, 100]$ and fixed quantization levels.
The received aggregate noise at agent $i$ is then
\begin{equation}
	e_i[t+1] = \sum_{j \in \mathcal{N}_i} w_{ij} ( \hat{x}_j[t]-x_j[t]).
\end{equation}
Since $\mathbf Q(\cdot)$ is unbiased with finite output range, $\{\mathbf e[t]\}$ is a MDS-type disturbance that satisfies Assumption \ref{assump.disturbance}.

For the stepsize sequences, we choose 
$$\eta[t] \!=\! 0.3 \left( 1 +0.9 t \right)^{-1} \text{, and  } \alpha[t] \!=\! 0.3 \left( 1 + 0.05\, t \right)^{-0.7},$$
which both satisfy Assumption \ref{Assumption: stepsize}.
Notably, the exponent $0.7$ is chosen so that $\alpha[t]$ decays sufficiently slowly to make the consensus evolution clearly visible in the simulation. Moreover, choosing the exponent $1$ for $\eta[t]$
gives $\eta[t]^2/\alpha[t]=\mathcal{O}(t^{-1.3})$, ensuring that \eqref{stepsizes relation} is satisfied.

\begin{figure}[h!]
	\centering
	\subfloat[PAC]{
		\includegraphics[width=0.385\textwidth]{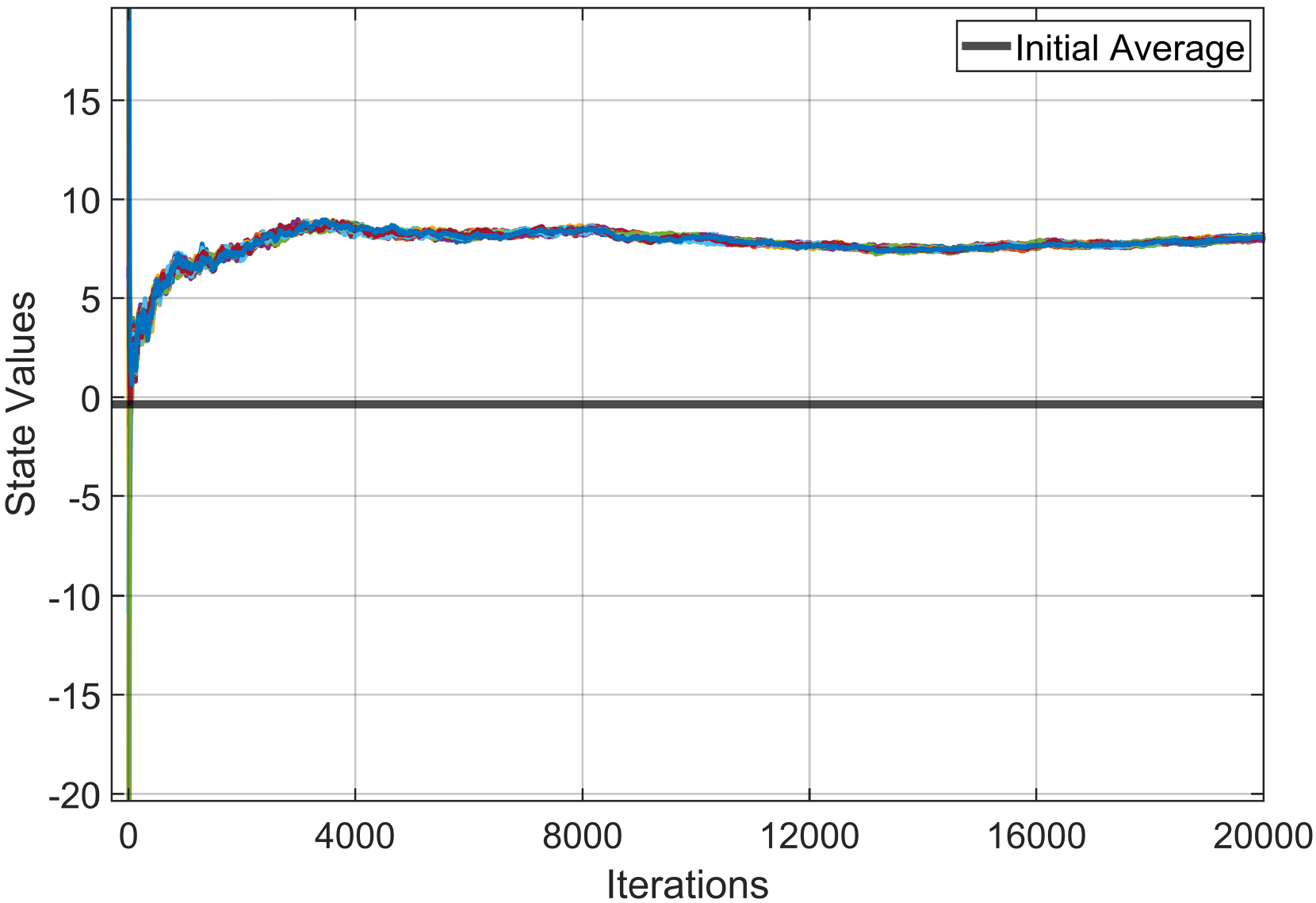}
		\label{Fig.1:sub1}
	}
	\hspace{0pt}
	\subfloat[AAC]{
		\includegraphics[width=0.385\textwidth]{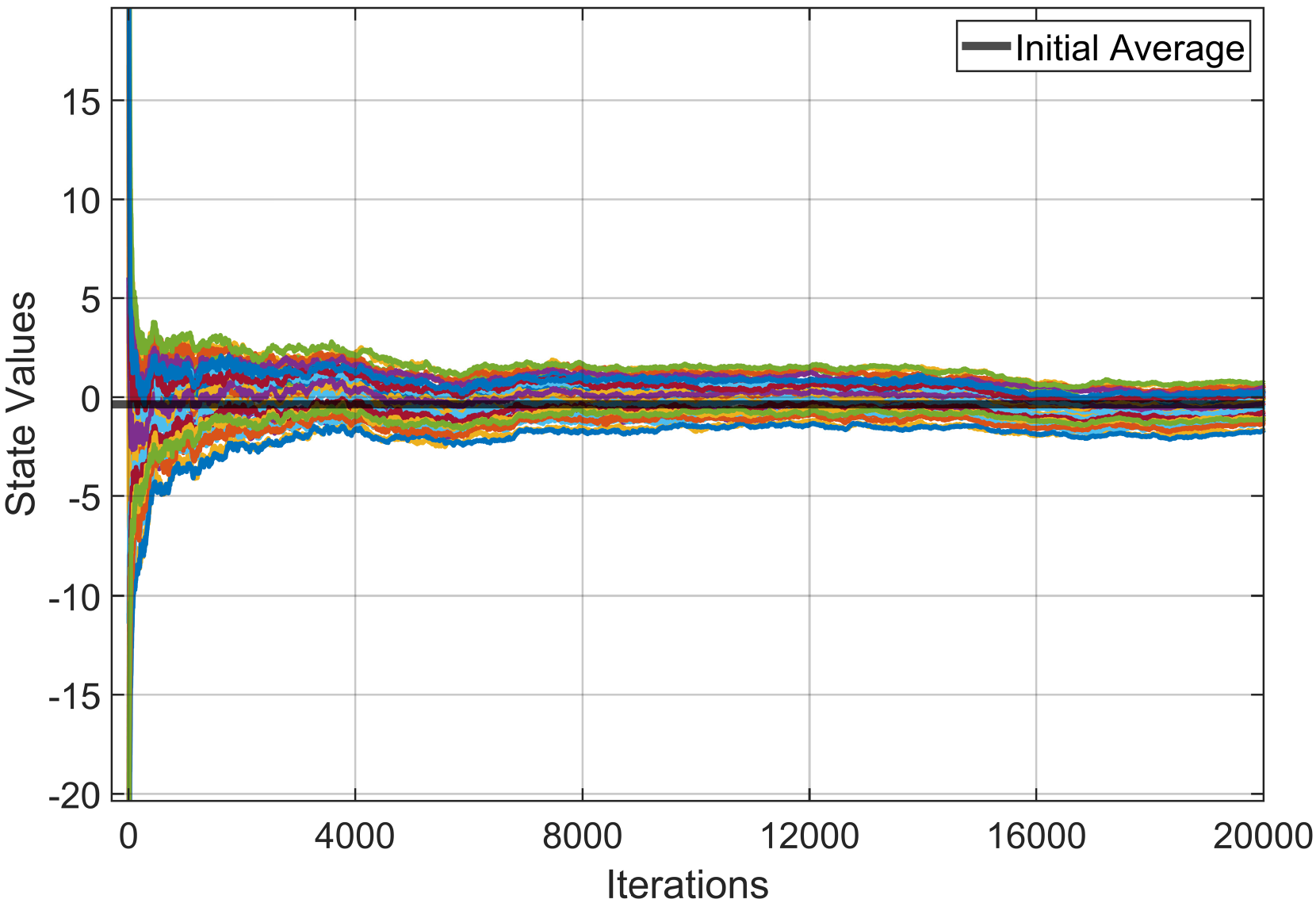}
		\label{Fig.1:sub3}
	}
	\caption{The state dynamics of $x_{n}[t]$ in $2\times10^4$ iterations.}
	\label{Fig.1}
\end{figure}

\begin{figure}[h!]
	\centering
	\includegraphics[width=0.385\textwidth]{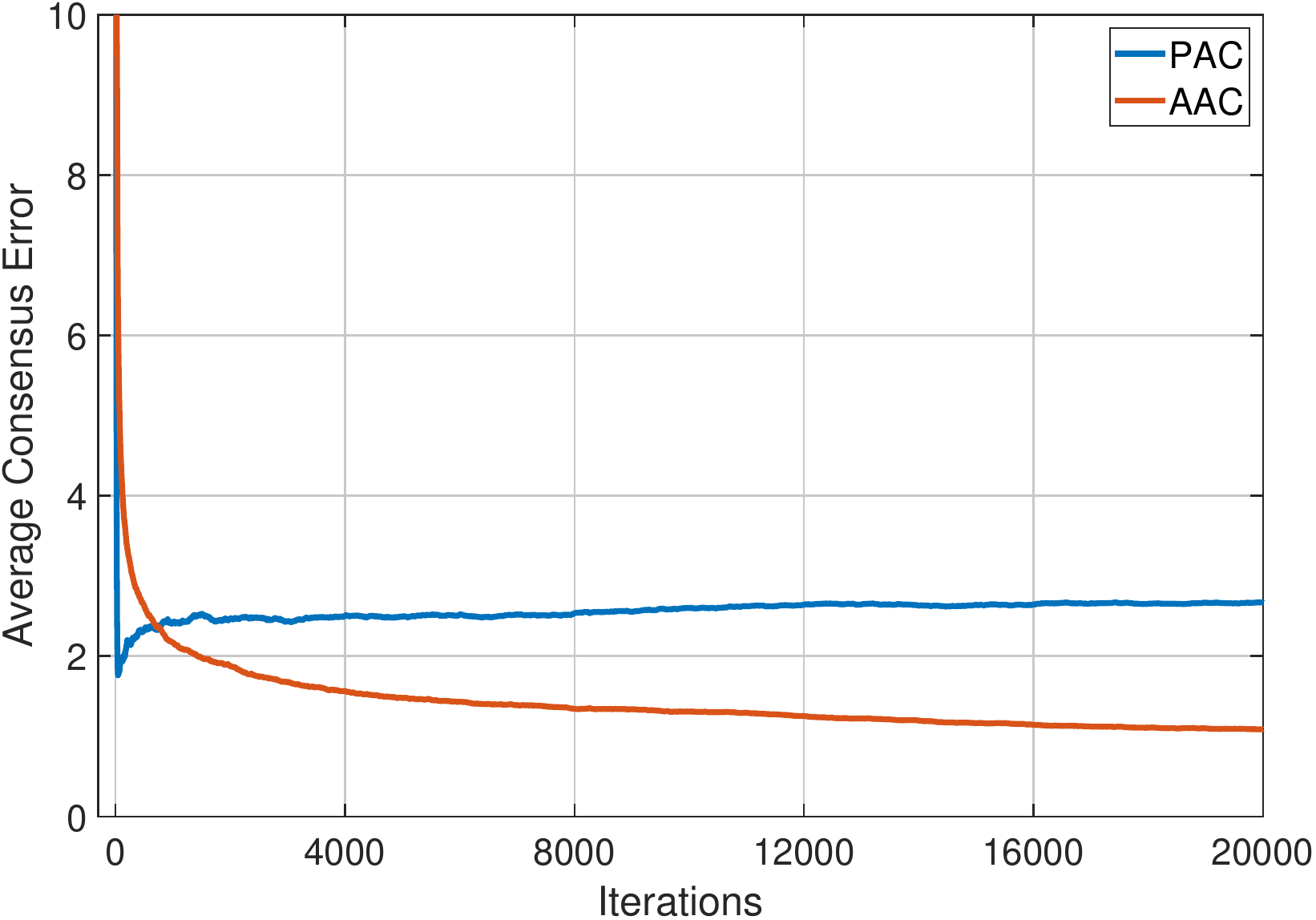}
	\caption{Convergence behavior.}
	\label{Fig.2}
	\vspace{-5mm}
\end{figure}

For performance evaluation, we adopt the average consensus error (ACE) $\psi [t]$ which is defined as
\begin{equation}
	\label{Eq.performance evaluation}	
	\psi[t] = \sqrt{\sum\nolimits_{n=1}^{N} (x_{n}[t] - x^{*})^{2}/N},
\end{equation}
where $x^* \!=\! \frac{1}{N} \sum_{i\in\mathcal{N}}x_i[0]$. We run both PAC and AAC for $2\times10^4$ iterations. The resulting state dynamics of $\mathbf x[t]$ in a single realization is presented in Fig. \ref{Fig.1}, while Fig. \ref{Fig.2} shows the ACE averaged over $100$ realizations.

Figures \ref{Fig.1} and \ref{Fig.2} show that AAC and PAC exhibit markedly different convergence behaviors under quantization noise. PAC drives the agent states to consensus rapidly, but the limiting consensus value deviates from the initial average $x^*$. Accordingly, its ACE $\psi[t]$ decreases quickly at the beginning and then stabilizes at a relatively high, nonzero value, indicating a steady-state bias caused by disturbance accumulation. This is consistent with the preceding analysis that PAC achieves consensus on an unbiased random variable with nonzero variance. 

By contrast, although the ACE of AAC decreases more slowly, it continues to decay towards zero, showing that AAC can effectively suppress disturbance accumulation and thereby eliminate the random bias. This improved robustness, however, comes at the expense of convergence speed, since the anchoring term continuously pulls each agent's state toward its initial value, thereby weakening the tendency to reach consensus rapidly. The less smooth AAC trajectories in Fig. \ref{Fig.1} also reflect the ongoing counteraction between the anchoring term and the random disturbances.

\section{CONCLUSIONS}
This paper studies the distributed average consensus problem under persistent link-level disturbances modeled as an MDS with uniformly bounded conditional second moments. Under such disturbances, standard SA-based average consensus algorithms can achieve consensus on a random variable whose expectation equals the initial network average, but generally cannot guarantee convergence to the exact average in each realization. 

Building upon a DGD recursion with SA techniques, we obtain a linear update rule that incorporates the initial state information into the average consensus iteration.
Since this term acts as an anchor that continuously pulls each agent state toward its initial value, we refer to the resulting scheme as anchor-based average consensus (AAC). We showed that this anchoring mechanism has an  optimization-based interpretation for counteracting disturbance accumulation, and that AAC achieves almost sure convergence to the exact average consensus value.

In summary, the main contribution of this paper is not the introduction of a completely new average consensus algorithm, but rather a principled interpretation and a rigorous convergence analysis of an anchoring-based noise-mitigating mechanism. This perspective also unifies several representative methods under a common viewpoint, including NR-PushSum and robust dynamic average consensus. The numerical results further support the analysis and highlight a trade-off: anchoring improves robustness and steady-state accuracy at the expense of convergence slowdown. Overall, both the theoretical and numerical results suggest that anchoring provides a useful and interpretable design for achieving exact average consensus under persistent disturbances.


\bibliographystyle{IEEEtran}
\bibliography{ref}
\end{document}